\documentclass[reprint,nofootinbib,amsmath,amssymb,aps,prmaterials]{revtex4-2}

\usepackage{graphicx}
\usepackage[dvipsnames]{xcolor}
\usepackage[colorlinks=true,
    linkcolor=blue,
    filecolor=blue,
    citecolor=blue,      
    urlcolor=blue,]{hyperref}

\usepackage{prettyref}
\newcommand{\pref}[1]{\prettyref{#1}}
\newrefformat{fig}{Fig.~\ref{#1}}
\newrefformat{tab}{Table~\ref{#1}}
\newrefformat{sec}{Sec.~\ref{#1}}
\newrefformat{ssec}{Sec.~\ref{#1}}
\newrefformat{app}{App.~\ref{#1}}
\newrefformat{eqn}{Eq.~(\ref{#1})}

\begin{document}

\title{Effects of strain on the stability of the metallic rutile and insulating M1 phases of vanadium dioxide}

\author{Peter Mlkvik}
\email{peter.mlkvik@mat.ethz.ch}
\author{Lena Geistlich}
\author{Nicola A. Spaldin}
\author{Claude Ederer}
\affiliation{Materials Theory, Department of Materials, ETH Z\"{u}rich, Wolfgang-Pauli-Strasse 27, 8093 Z\"{u}rich, Switzerland}

\date{\today}

\begin{abstract}
We present a systematic density-functional theory study of the effects of strain on the structural and electronic properties in vanadium dioxide (VO$_2$), with particular emphasis on its effect on the relative stability of the metallic rutile and the insulating monoclinic M1 phases. We consider various strain conditions that can be related to epitaxial strain present in VO$_2$ films grown on different lattice planes. Our calculations confirm the dominant role of $c$ axis strain, i.e., along the direction of the V--V dimerization in the M1 phase. Our analysis suggests that this effect stems primarily from the weakening of the lattice stiffness, with the hopping along the $c$ axis playing a minor role. We also confirm that, in strain scenarios that deform the basal plane, the $c$ axis strain still has a dominant effect on the phase stability.
\end{abstract}

\maketitle

%%%%%%%%%%%%%%%%%%%%%%%%%%%%%%%%%%%%%%%%%%%%%%%%%%%%%%%%%%%%%%%%%%%%%%%%%%%%%%%%

\section{Introduction}

Vanadium dioxide (VO$_2$) is a prototypical system undergoing a metal-insulator transition (MIT). The proximity of the MIT temperature, $T_c$, to room temperature makes VO$_2$ a particularly promising material for a wide range of potential applications~\cite{Yang/Ko/Ramanathan:2011, Liu_et_al:2018}. Although the tunability of the MIT through external stimuli such as doping or strain~\cite{Cao_et_al:2009, Shao_et_al:2018, Shi_et_al:2019} has been demonstrated, a comprehensive understanding of the underlying mechanisms is lacking. Here, we present a systematic computational study of the electronic and structural effects of strain on the energy difference between two principal phases of VO$_2$.

\begin{figure}
        \centering
	\includegraphics[width=0.9\linewidth]{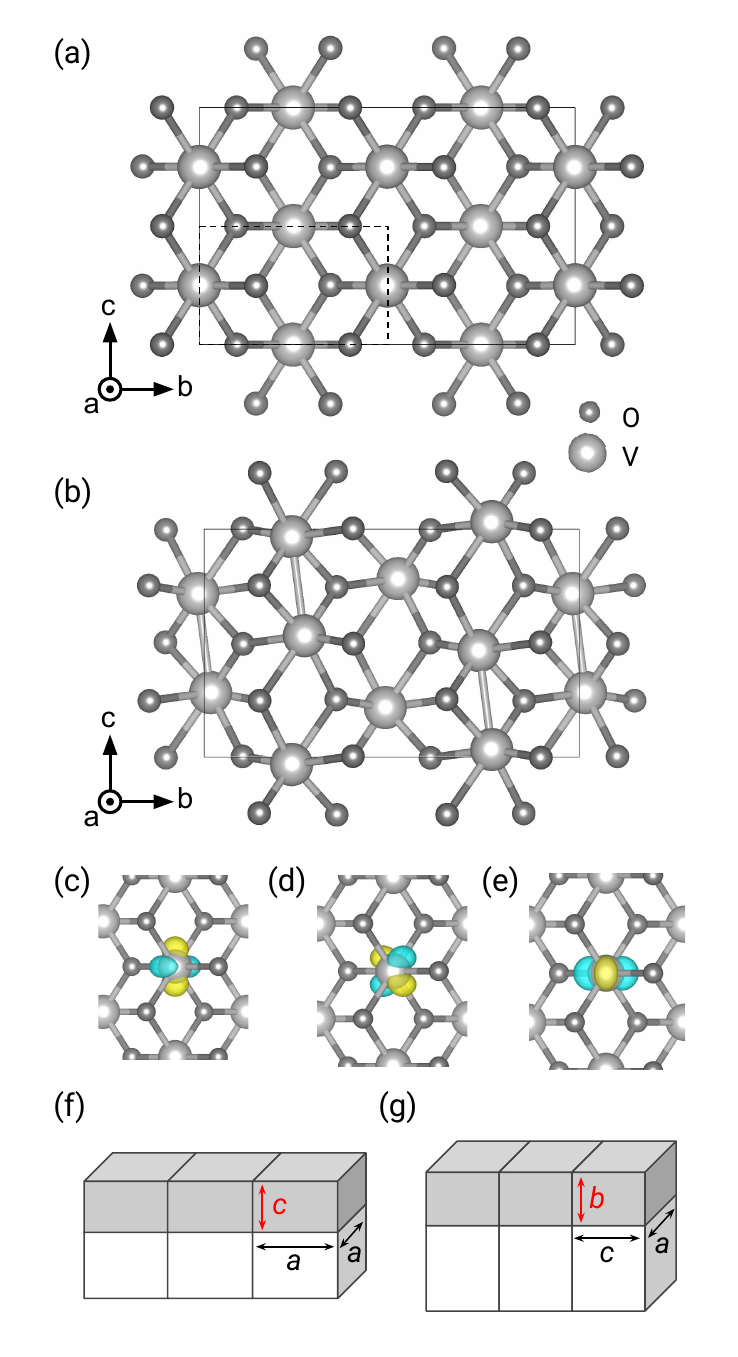}
	\caption{(a, b) Supercell used in the calculations to describe both the (a) R and (b) M1 structures. The dashed lines in (a) indicate the underlying primitive rutile cell. Arrows indicate the definition and orientation of the corresponding orthorhombic axes. V (O) atoms shown in light (dark) gray. (c-e) The three V $t_{2g}$ orbitals, $d_{x^2-y^2}$, $d_{xz}$, and $d_{yz}$, respectively, highlighting their orientation with respect to the local environment. Yellow (cyan) colors indicate positive (negative) phase of the orbitals. (f, g) The two epitaxial strain orientations considered in this work, corresponding to growth along the (f) [001] and (g) [010] directions. Axes shown in red are free to relax, while the axes shown in black are fixed by the substrate.}
	\label{fig:structure}
\end{figure}

During the MIT, chains of vanadium dimers form along the $c$ direction of the underlying high-temperature metallic rutile (R) structure, resulting in the low-temperature insulating monoclinic M1 structure~\cite{Eyert:2002a}~[see \pref{fig:structure}(a) and \pref{fig:structure}(b), respectively]. The main physics behind this transition was first suggested by Goodenough~\cite{Goodenough:1971}, and assumes a singlet formation within the vanadium--vanadium dimers, due to the pairing of the electrons occupying the lowest lying $d$ orbital of the nominal $d^1$ vanadium cations. Defining a local coordinate system with the local $x$ axis along $c$ and the local $z$ axis along the apical vanadium-oxygen bond~\cite{Eyert:2002a}, this lowest lying orbital is the $d_{x^2-y^2}$ orbital [\pref{fig:structure}(c)]. The large bonding-antibonding splitting resulting from the vanadium-vanadium dimerization together with a depletion of the $d_{xz}$ and $d_{yz}$ orbitals [\pref{fig:structure}(d, e)] associated with a zig-zag distortion perpendicular to $c$, gives rise to insulating behavior. The simplicity of this picture has, however, been called into question in part due to the presence of a Mott-insulating monoclinic M2 phase~\cite{Pouget_et_al:1975}, and thus the exact origin of the electronic and structural transition in VO$_2$ has been a topic of controversy for decades~\cite{Zylbersztejn/Mott:1975, Wentzcovitch/Schulz/Allen:1994, Rice/Launois/Pouget:1994}. Currently, the consensus characterizes the material as a correlation-assisted Mott-Peierls insulator~\cite{Biermann_et_al:2005, Weber_et_al:2012, Brito_et_al:2016}. 

Large changes to the MIT temperature, $T_c$, have been observed in VO$_2$ samples under strain. Strain has been applied to VO$_2$ through bending~\cite{Cao_et_al:2010, Atkin_et_al:2012}, pressure~\cite{Chen_et_al:2017a, Birkholzer_et_al:2022}, nanowire engineering~\cite{Zhang/Chou/Lauhon:2009, Cao_et_al:2009, Park_et_al:2013, Asayesh-Ardakani_et_al:2015}, or through growth of thin films on substrates with a certain lattice mismatch and different surface orientations~\cite{Muraoka/Hiroi:2002, Yang_et_al:2010, Aetukuri_et_al:2013,Quackenbush_et_al:2013a, Quackenbush_et_al:2016, Lee_et_al:2017, Fischer_et_al:2020, Lee_et_al:2022}, giving rise to coherent epitaxial strain along different crystallographic directions. In particular, it has been shown that growth with the rutile $c$ axis oriented either perpendicular to [\pref{fig:structure}(f)] or within [\pref{fig:structure}(g)] the substrate surface plane of a TiO$_2$ substrate suppresses and enhances $T_c$, respectively~\cite{Muraoka/Hiroi:2002, Quackenbush_et_al:2016}. This observation has been rationalized by the different orientation and sign of the strain relative to the direction of dimerization, which is strongly linked to the hopping between the Peierls-active $d_{x^2-y^2}$ orbitals~\cite{Muraoka/Hiroi:2002}. It has thus been suggested that elongating $c$ leads to an increase in $T_c$, while compressing $c$ decreases $T_c$. However, the role of strain along other crystallographic directions is still debated~\cite{Quackenbush_et_al:2016}. For example, it has been suggested that, as the basal lattice parameters contract, the relative oxygen apical and equatorial distances change, and the filling of the $d_{x^2-y^2}$ orbital is affected~\cite{Aetukuri_et_al:2013, Lee_et_al:2022}. Computationally, the importance of the bandwidth change and the corresponding change in hopping amplitudes along the dimerization direction have been highlighted~\cite{Lazarovits_et_al:2010}. However, a systematic study detailing both structural and electronic effects for the two relevant VO$_2$ phases under a wide range of strain conditions is lacking.

In this work, we provide such a systematic first-principles study of different strain conditions and their effects on VO$_2$. In particular, we investigate how the interplay between electronic and structural degrees of freedom determines the energetic stability of the R and M1 phases. We focus on coherent strain corresponding to epitaxial growth on the (001) plane of the rutile structure, but we also exploit the flexibility of our computational approach to vary the in-plane and out-of-plane strains independently from each other. Finally, we also briefly explore the effect of applying different strains along the two orthogonal directions within the basal plane, corresponding, e.g., to epitaxial growth on the (010) plane, which breaks the tetragonal symmetry of the underlying rutile structure. We confirm that the VO$_2$ phase stability is primarily governed by the strain along the dimerization direction, which modifies the distance between nearest-neighbor vanadium atoms and, consequently, affects both the hopping amplitudes and the structural stiffness.

%%%%%%%%%%%%%%%%%%%%%%%%%%%%%%%%%%%%%%%%%%%%%%%%%%%%%%%%%%%%%%%%%%%%%%%%%%%%%%%%

\section{Methodology}

To simulate different strain conditions in VO$_2$, we embed both the R and M1 structures in the same orthorhombic unit cell formed from four primitive cells of the underlying rutile structure, corresponding to a $1 \times 2 \times 2$ supercell [see \pref{fig:structure}(a) and (b), with the primitive rutile cell indicated by dashed lines in (a)]. We use an orthorhombic cell for both structures (i.e., without monoclinic strain in the M1 structure) in order to allow for a more systematic comparison between the different phases. We define $c$ along the dimerization direction, aligning $a, b,c$ with $x, y,z$ of the global coordinate system.

In \pref{sec:epitaxial}, we consider only strain conserving the tetragonal symmetry of the R phase, i.e., we keep the $a:b$ ratio of the supercell lattice vectors fixed to $1:2$, applying $\epsilon_{xx}=\epsilon_{yy}$ strain, and observing the change in $c$ by relaxing the $\sigma_{zz}$ stress component. In \pref{sec:map}, we also keep the tetragonal symmetry but now vary both $\epsilon_{xx}=\epsilon_{yy}$ and $\epsilon_{zz}$ strains independently. Finally, in \pref{sec:orthorhombic}, we investigate the orthorhombic strain regime by imposing $\epsilon_{xx} \neq \epsilon_{yy}$ and $\epsilon_{zz}$.

We define the applied strain as $\epsilon_{xx}=(a-a_0)/a_0 \times 100\%$, where $a$ is the strained lattice parameter and $a_0=4.630$\,Å is its unstrained reference defined by the fully relaxed rutile R structure within DFT$+V$ and using the Perdew-Burke-Ernzerhof (PBE) exchange-correlation functional~\cite{Perdew/Burke/Ernzerhof:1996}, so that the definition of strain is not affected by the typical small over- and underestimation of the $a$ and $c$ lattice parameters within PBE compared to experiment. For each strain, we perform relaxations of the internal positions of all atoms starting from both R and M1 internal positions, and then observe the changes in energy as well as electronic and structural properties.

To account for the strong intersite effects along the dimerizing vanadium chains, we utilize the DFT$+V$ machinery as presented in Ref.~\cite{Haas_et_al:2024}. We employ the $+V$ correction as an empirical parameter, accounting for the intersite static self-energy necessary to enhance dimerization in the M1 phase~\cite{Tomczak/Biermann:2007}. In line with our previous findings~\cite{Haas_et_al:2024}, and unless otherwise noted, we fix $V=2$\,eV on both short and long bonds (SB and LB), obtaining the correct energetic ordering of the unstrained R and M1 phases, and the correct size of the band gap in the M1 phase. 

We perform the DFT$+V$ calculations using the \textsc{quantum espresso} (v7.3) package~\cite{Giannozzi_et_al:2009, Giannozzi_et_al:2017} within the generalized gradient approximation using the PBE functional~\cite{Perdew/Burke/Ernzerhof:1996}. We use the ultrasoft pseudopotentials from the GBRV library~\cite{Garrity_et_al:2014}, including the 3$s$ and $3p$ semicore states in the valence manifold of the vanadium atoms. We use a plane-wave kinetic energy cutoff of 70~Ry and 12$\times$70\,Ry for the charge density. We use a $\Gamma$-centered 8$\times$5$\times$6 $k$-mesh, and converge the change in total energies between consecutive iterations to be smaller than 5$\times$10$^{-8}$\,eV, and all force components to be smaller than $10^{-3}$\,eV/Å.

We construct a localized basis set using \textsc{wannier90}~(v3.1.0)~\cite{Mostofi_et_al:2014, Pizzi_et_al:2020}. We include the set of bands around the Fermi level to encompass the whole \{$d_{x^2-y^2}$, $d_{xz}$, $d_{yz}$\} manifold, and the resulting Wannier bands closely reproduce the underlying DFT band structure~(for the DFT and Wannier bands, see, e.g., Fig.~2 in Ref.~\cite{Mlkvik_et_al:2024}). We use this basis to extract the hopping amplitudes between all pairs of orbitals and their respective on-site energies.

%%%%%%%%%%%%%%%%%%%%%%%%%%%%%%%%%%%%%%%%%%%%%%%%%%%%%%%%%%%%%%%%%%%%%%%%%%%%%%%%

\section{Results and Discussion}

\subsection{(001)-oriented epitaxial strain}
\label{sec:epitaxial}

We first discuss the behavior of the R and M1 phases in VO$_2$ under a strain of $\epsilon_{xx}=\epsilon_{yy}$, i.e., maintaining the quasi-tetragonal symmetry of the supercell lattice vectors, $b=2a$, and allowing the cell to relax along the $c$ axis, mimicking the epitaxial strain conditions of a thin film grown on a (001)-oriented square-lattice substrate [see \pref{fig:structure}(f)]. This orients the vanadium chains along the growth direction, perpendicular to the in-plane orientation of the film.

\begin{figure}
        \centering
	\includegraphics[width=0.9\linewidth]{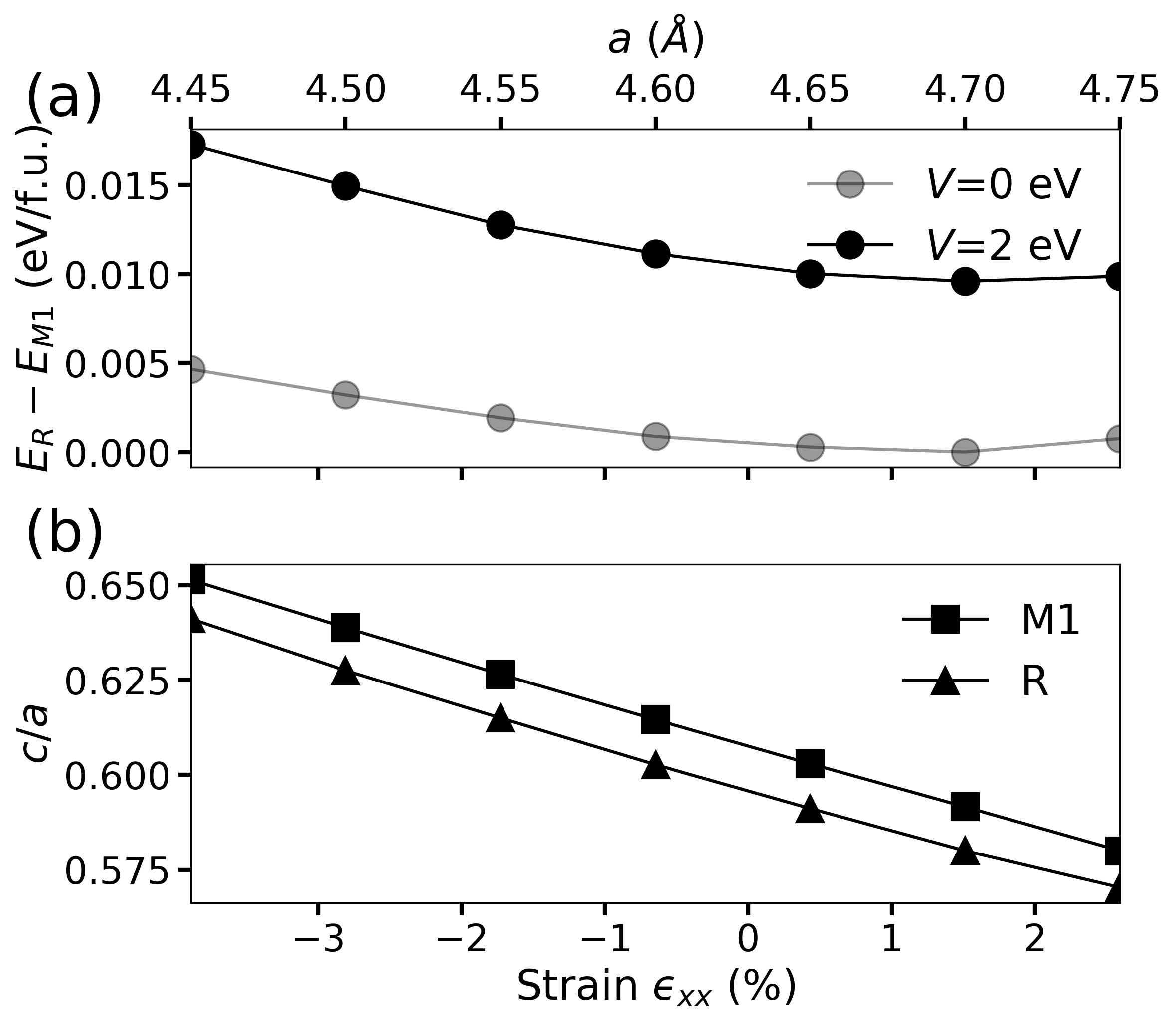}
	\caption{(a) Energy difference between the R and M1 structures as a function of strain, $\epsilon_{xx}=\epsilon_{yy}$, calculated with (without) the $+V$ correction in black (gray). (b) The $c/a$ ratio as a function of strain for the R and M1 phases marked by triangles and squares, respectively. Strain is defined with respect to the relaxed R phase lattice parameter, $a_0=4.630$\,Å.}
	\label{fig:astrain}
\end{figure}

In \pref{fig:astrain}(a), we show the difference in total energies of both phases as a function of epitaxial strain. We find that the M1 phase is lower in energy than R for all strains, with the energy difference increasing under compressive and decreasing under moderate tensile strain, with a minimal energy difference at around $\epsilon_{xx}=2\%$. We also verify this result using standard DFT calculations without an intersite correction, i.e., $V=0$\,eV, for which the energy difference between the two phases is significantly smaller but shows an identical trend with strain. 

Indeed, if we interpret the relative phase stability as a proxy for the transition temperature, $T_c$, our results imply that increasing the basal plane lattice constant $a$ decreases the stability of the M1 phase, matching the experimentally observed reduction in $T_c$ under tensile epitaxial strain observed in thin films grown along (001)~\cite{Muraoka/Hiroi:2002, Aetukuri_et_al:2013, Quackenbush_et_al:2013a}. We discuss this effect in a more comprehensive picture in \pref{sec:map}.

Additionally, as shown in \pref{fig:astrain}(b), increasing the in-plane lattice parameter $a$ leads to a shortening along $c$. Notably, the M1 phase always exhibits a larger $c/a$ ratio than the R phase. Therefore, naturally, this type of strain results in an interplay of two effects; not only the direct impact of the in-plane lattice parameter $a$, but also the shortening of $c$ due to the Poisson effect. Previously, it has been suggested that it is the latter that is the main factor determining the change in $T_c$~\cite{Muraoka/Hiroi:2002, Quackenbush_et_al:2013a}. In the following, we hence consider these two strains independently.

\subsection{General tetragonal strain}
\label{sec:map}

\begin{figure}
        \centering
	\includegraphics[width=1\linewidth]{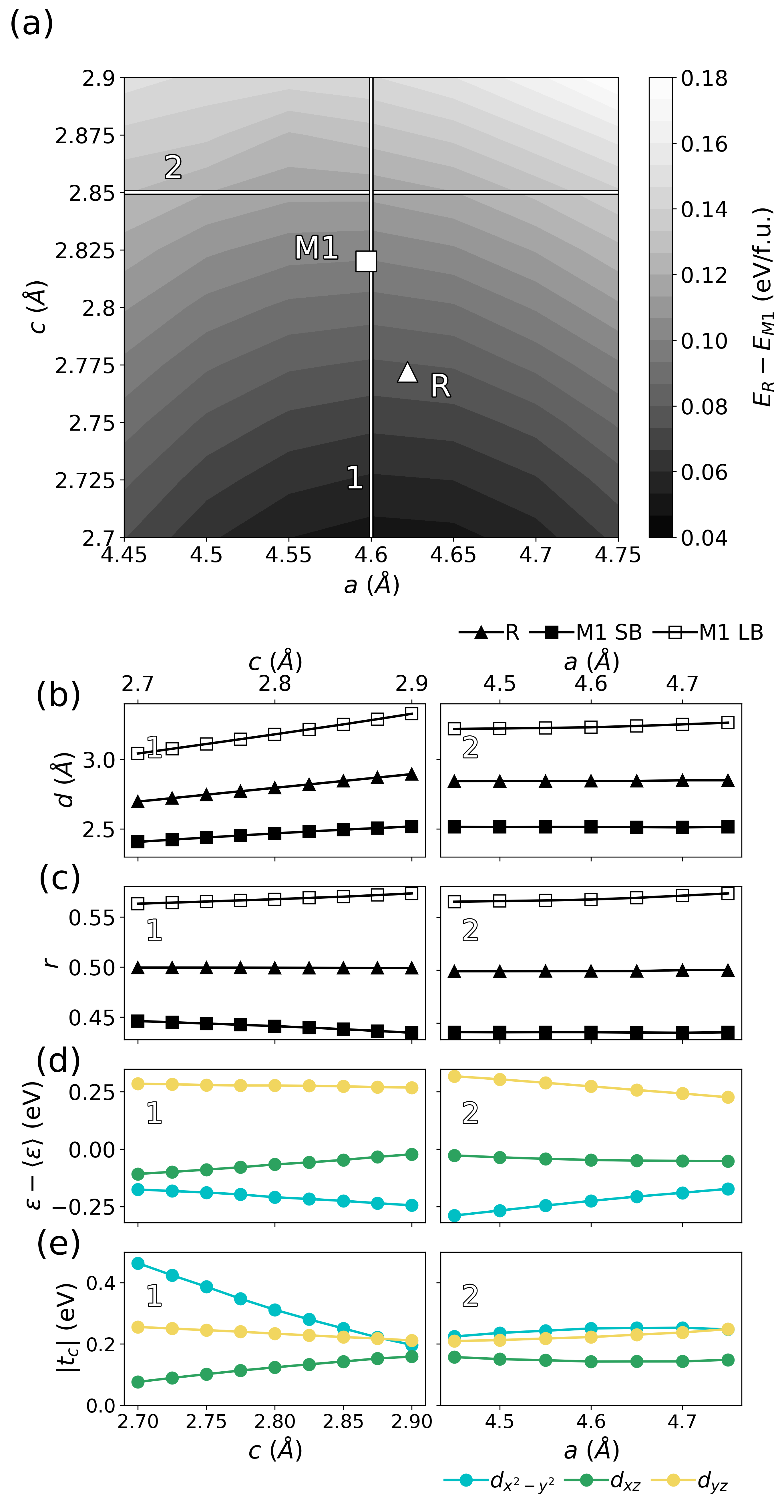}
	\caption{(a) Energy difference between the R and M1 phases. Triangle (square) indicates the relaxed R (M1) lattice parameters. Numbered lines indicate cuts along different directions. (b) Nearest-neighbor V--V distance, $d$, as a function of $c$ (Cut 1, left) or $a$ (Cut 2, right) lattice parameter, and (c) ratio of the nearest-neighbor V--V distance to $c$, $r=d/c$. Triangular markers indicate R and filled (empty) square markers indicate M1 SB (LB). (d) Energy levels, $\varepsilon$, relative to the average energy level, $\langle \varepsilon \rangle$, and (e) nearest neighbor hopping amplitudes along $c$, $t_c$, both corresponding to $d_{x^2-y^2}$ (cyan), $d_{xz}$ (green), $d_{yz}$ (yellow) orbitals,  as function of $a$ or $c$ in the R phase.}
	\label{fig:map}
\end{figure}

To facilitate a deeper insight into the effect of epitaxial strain and to disentangle the effects caused by changes in different lattice directions, we now discuss the behavior of the R and M1 phases in VO$_2$ under a general tetragonal strain (i.e., with $b=2a$) but treating the $a$ and $c$ lattice parameters as independent variables (i.e., independently applying $\epsilon_{xx}=\epsilon_{yy}$ and $\epsilon_{zz}$). 

In \pref{fig:map}(a), we show the energy difference between the R and M1 phases as a function of $a$ and $c$. We show a range of values around the equilibrium $a$ and $c$ of the R and M1 phases [\pref{fig:map}(a), triangular and square markers, respectively]. In this whole strain range, for this choice of exchange-correlation functional and $V$ value, the M1 phase is always lower in energy compared to R. With increasing $c$, the M1 phase becomes more energetically favorable for all $a$ [\pref{fig:map}(a), lighter color for increasing $c$]. This is consistent with the observation of a larger $c$ lattice parameter of the M1 phase compared to the R phase [as seen in \pref{fig:astrain}(b) and indicated by the markers in \pref{fig:map}(a)]. Notably, we also observe the M1 phase becoming more favorable at both larger and smaller values of $a$ than its equilibrium value [\pref{fig:map}(a), lighter color at both large and small $a$ values], although this effect appears weaker compared to the dependence on $c$.

To understand the origin of these trends, we now examine two different cuts through the $a$-$c$ diagram shown in \pref{fig:map}(a) for fixed values of $a=4.6$\,Å and $c=2.85$\,Å, respectively [Cuts 1 and 2 in \pref{fig:map}(a)], and analyze the corresponding changes in structural and electronic properties in more detail. We first consider the nearest-neighbor V--V distance [\pref{fig:map}(b)] and the ratios of these distances to $c$ [\pref{fig:map}(c)]. 

\begin{figure}
        \centering
	\includegraphics[width=0.9\linewidth]{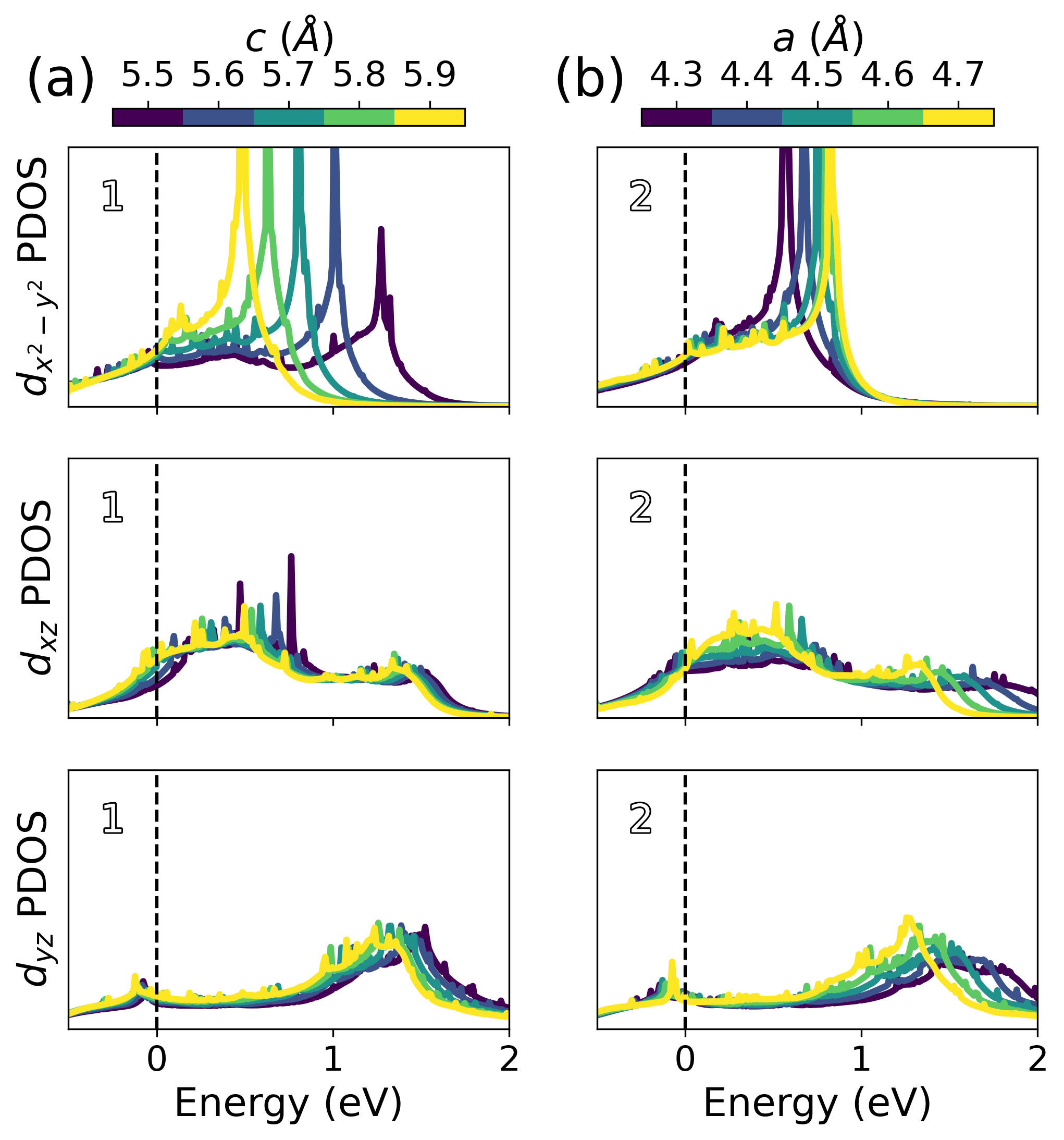}
	\caption{Evolution of the PDOS of the different $t_{2g}$ orbitals within the R phase (a) along Cut 1 and (b) along Cut 2. Color gradient indicates increasing lattice parameter. The vertical dashed line indicates the Fermi level.}
	\label{fig:pdos}
\end{figure}

Along Cut 1, i.e., variation along the dimerization axis, $c$, VO$_2$ shows an increase in all nearest neighbor V--V distances, also for the SB distance in M1 [\pref{fig:map}(b) left, all lines increasing as $c$ increases]. However, although the absolute V--V distances increase, the M1 SB bond length decreases relative to $c$, i.e., $r_\text{SB}=d_\text{SB}/c$ decreases [\pref{fig:map}(c) left, square markers trending down as $c$ increases], suggesting a strengthening of dimerization with increasing $c$.

Along Cut 2, i.e., basal plane strain at fixed $c$, the effects are more subtle. The nearest neighbor distance remains largely unchanged [\pref{fig:map}(b) right, all lines remain constant], except for a very weak upwards trend in $d_\text{LB}$ and $r_\text{LB}$ related to an increase of the zig-zag distortion (not shown). This is caused by the expansion in the basal plane and affects mainly the long bond.

To rationalize the energetic competition between the R and M1 phases, we now draw on a simplified Peierls-like model, where a periodic structural displacement modulates the hopping amplitude between neighboring sites, leading to an energy gap and a reduction in total energy. The key ingredients thereby are the hopping amplitude between neighboring atoms and the stiffness of the underlying crystal lattice. In VO$_2$, the quasi-1D character of the $d_{x^2-y^2}$ orbital along the $c$ axis makes this framework a useful approximation. 

Whether the distortion becomes energetically favorable depends on the balance between two effects~\cite{Su/Schrieffer/Heeger:1980}. An energy gain arises from electronic stabilization due to gap formation related to the difference in the hopping amplitudes of the SB and LB, which can be related to the slope of the hopping with respect to interatomic distances. On the other hand, the distortion also leads to an energy cost that depends on 
%an elastic penalty due to 
the stiffness of the phonon mode related to the structural dimerization. Furthermore, in the multiband case realized in VO$_2$, the small deviation from half-filling of the dimerizing $d_{x^2-y^2}$ orbital is expected to be relevant. To gain insight into these effects, we track the energy levels of the different $t_{2g}$ orbitals [\pref{fig:map}(d)], the corresponding hopping amplitudes, $t_c$, between nearest neighbors along the dimerization direction [\pref{fig:map}(e)], and the projected densities of states (PDOS) (\pref{fig:pdos}), all in the undimerized R phase.

Along Cut 1, the increasing nearest-neighbor V--V distances cause an energy lowering of the $d_{x^2-y^2}$ orbital compared to the other two orbitals [\pref{fig:map}(d) left, cyan line coming down in energy]. 
This can be rationalized by the fact that the $d_{x^2-y^2}$ orbital points directly in between the oxygen ligands along $c$, and thus moving these ligands further away by increasing $c$ is expected to lower its energy. A similar, albeit much weaker, energy lowering can also be expected for the $d_{xz}$ orbital, which can appear as an increase if taken relative to the average orbital energy. The lowering of the $d_{x^2-y^2}$ orbital can support the depletion of the other two orbitals in favor of a half-filling of the $d_{x^2-y^2}$ orbital, and thus further stabilize the M1 phase, even though the PDOS of the R phase [\pref{fig:pdos}(a) left] does not indicate any noticeable changes in the orbital fillings for increasing $c$. On the other hand, the hopping amplitude between the $d_{x^2-y^2}$ orbitals along $c$ decreases strongly with increasing $c$ [\pref{fig:map}(c) left]. This effect is also visible as a narrowing of the corresponding PDOS [\pref{fig:pdos}(a) left]. If one makes the reasonable assumption that a smaller hopping amplitude also results in a smaller slope with respect to the V--V distance (e.g., if one assumes a typical power-law dependence), one would in fact expect a smaller electronic energy gain from the dimerization at larger $c$, and thus a disfavoring of the M1 phase. However, the decrease of the hopping with $c$ in \pref{fig:map}(c) appears close to linear, and thus its slope remains nearly constant, suggesting that this effect might be weak. Furthermore, an increase in $c$ is also expected to weaken the inter-ionic forces and thus to reduce the elastic stiffness opposing the structural dimerization. We thus conclude that the observed stabilization of the M1 phase for an elongated $c$ lattice parameter results from the interplay between different electronic and structural effects, with the softening of the structural stiffness potentially being the decisive factor.

Along Cut 2, changing the basal plane lattice parameter $a$ for constant $c$ also shifts the energy of the $d_{x^2-y^2}$ orbital relative to the other two orbitals, but now the $d_{x^2-y^2}$ orbital is lowered for decreasing $a$ [\pref{fig:map}(d) right, cyan line]. Regarding the hopping amplitudes, while the hoppings in the basal plane (not shown) show the expected behavior (increasing with decreasing $a$), there is only a very weak effect on the hoppings along $c$, even though we observe a subtle decrease in $t_c$ corresponding to the $d_{x^2-y^2}$ orbital for small $a$ [\pref{fig:map}(c) right, cyan line]. A potential explanation for the stabilization of the M1 phase for small $a$ could therefore be related to the lowering of the $d_{x^2-y^2}$ orbital, while the stabilization for large $a$ could again be due to a general softening of the structural stiffness, in this case affecting mostly the zig-zag component of the structural distortion. However, these trends are weak and thus also more subtle effects not considered here could come into play.

Based on this analysis, we can now also interpret the epitaxial strain results discussed in \pref{sec:epitaxial}. Under tensile strain, we observe that the increase in $a$ is accompanied by a contraction in $c$ [\pref{fig:astrain}(b)]. This shortening of the $c$ axis increases the hopping amplitude, which could, in principle, favor the Peierls distortion, but it also increases the stiffness opposing the Peierls distortion. The stabilization of the M1 phase under $c$ axis compression is thus governed by the interplay of two opposing tendencies, whereby the softening of the stiffness appears to dominate. Furthermore, the observed reduction in the energy difference between R and M1 under tensile strain primarily reflects the effect of $c$ axis compression rather than the in-plane expansion, even though both effects are interdependent.

\subsection{Orthorhombic strain}
\label{sec:orthorhombic}

\begin{figure}
	\centering
	\includegraphics[width=0.9\linewidth]{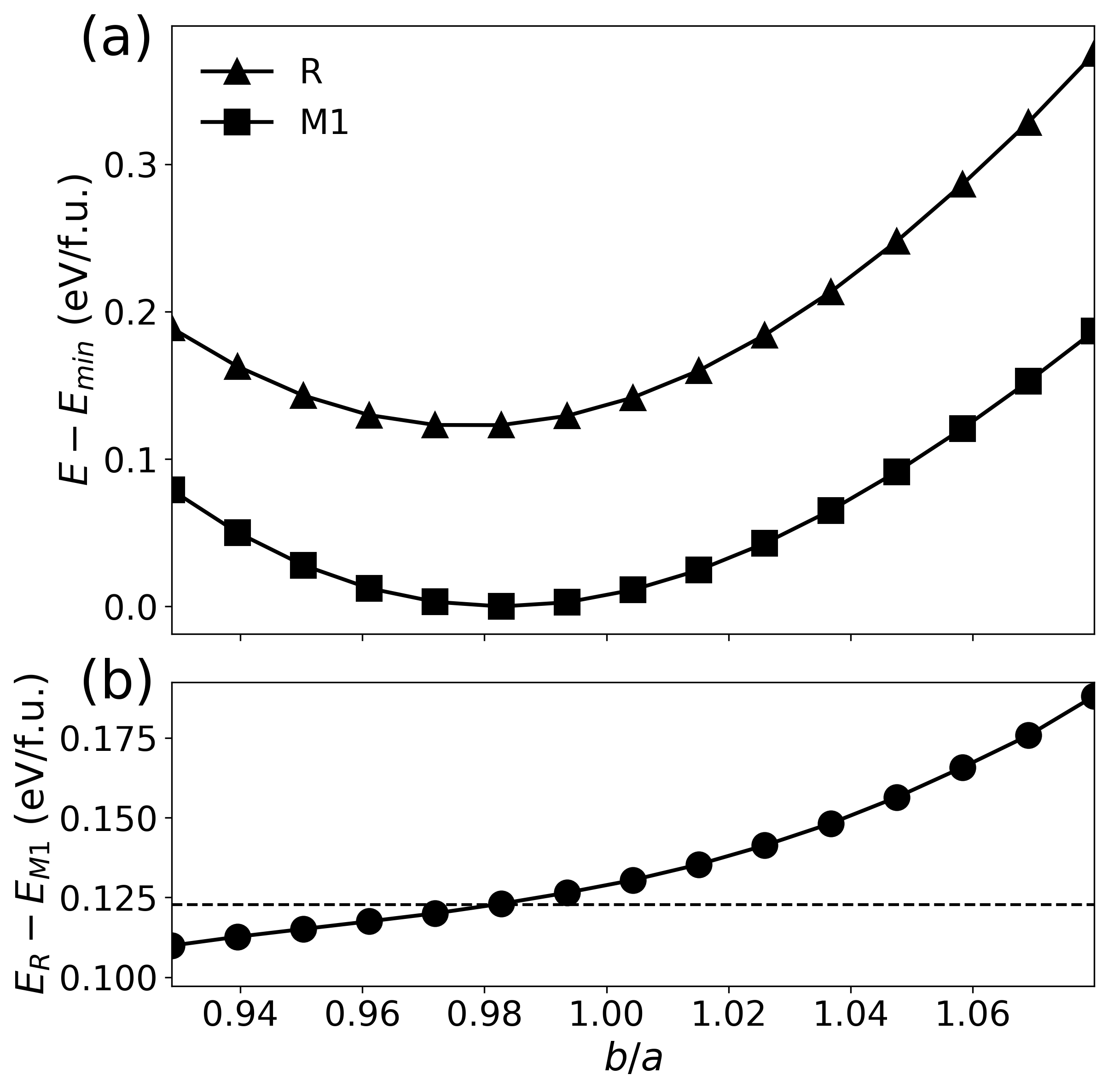}
	\caption{(a) Energies of the R and M1 phases as a function of the $b/a$ ratio, indicated by triangular and square markers, respectively, with $a$ and $c$ strained to values corresponding to a (010)-oriented TiO$_2$ substrate. (b) Energy difference between the two phases, with the dashed line indicating the corresponding energy difference at $c=c(\text{TiO}_2)$ and the corresponding equilibrium $a$. }
	\label{fig:bstrain}
\end{figure}

Finally, we also consider a strain condition that breaks the tetragonal symmetry of the in-plane lattice parameters, corresponding to VO$_2$ epitaxially grown with (010) orientation, i.e., with the vanadium dimerization axis lying in-plane [see \pref{fig:structure}(g)]. This approach complements our previous results by generalizing them to a less symmetric scenario with $\epsilon_{xx} \neq \epsilon_{yy}$ and $\epsilon_{zz}$, aiming to verify the simplified picture above.

To make this strain setup experimentally relevant, we perform the analysis at the strain values imposed by a (010)-oriented TiO$_2$ substrate, a commonly used rutile substrate for VO$_2$~\cite{Muraoka/Hiroi:2002, Quackenbush_et_al:2016}. Thus, we fix the $a$ and $c$ lattice parameters of our supercell to strains of 0.863\% and 3.62\%, respectively, corresponding to the nominal lattice mismatch of VO$_2$ grown on the (010) surface of TiO$_2$~\cite{Muraoka/Hiroi:2002}, and we then vary the $b$ lattice parameter.

In \pref{fig:bstrain}(a), we show the energy of the R and M1 phases as a function of $b/a$ in the above-described strain regime. Due to the tensile strain in both $c$ and $a$, both curves feature a minimum at $b/a<1$. The positions of the two minima almost coincide around $b/a=0.98$, with the R phase minimum located at a slightly smaller $b/a$ as that of the M1. In \pref{fig:bstrain}(b), we also show the energy difference between the two structures as a function of $b/a$. Similar to what we observed previously for tetragonal symmetry, the M1 phase is always lower in energy. 

To quantify the effect of the symmetry breaking due to the different strain values along $a$ and $b$, we also include the energy difference between R and M1 obtained under the tetragonal strain condition from \pref{sec:map} taken at the same $c$ axis strain of 3.62\% and the values of $a$ that minimize the total energy for that $c$ (which are nearly identical for R and M1). We observe that this energy difference [dashed line in \pref{fig:bstrain}(b)] matches almost exactly that of the orthorhombic strain at $b/a=0.98$ [dashed line crosses black line at exactly 0.98 in \pref{fig:bstrain}(b)]. This indicates that even in the lower orthorhombic symmetry, with different strains along the two basal plane directions, the energy difference between R and M1 is mainly determined by the $c$ axis strain, validating the assumptions made in, e.g., Refs.~\cite{Muraoka/Hiroi:2002, Quackenbush_et_al:2016}.

%%%%%%%%%%%%%%%%%%%%%%%%%%%%%%%%%%%%%%%%%%%%%%%%%%%%%%%%%%%%%%%%%%%%%%%%%%%%%%%%

\section{Summary and Outlook}

In this work, we have presented a systematic study of how strain affects the relative energetic stability of the metallic R and insulating M1 phases of VO$_2$, considering both structural and electronic effects arising from different strain geometries, including strain along the dimerization direction as well as strain within the basal plane. 

We first examine the effects of coherent epitaxy within the (001) plane. We show that increasing in-plane strain leads to a reduction of the $c/a$ ratio, and a decreased stability of the M1 phase. This behavior, driven predominantly by the shortening of the $c$ axis, reflects the suppression of the Peierls-distorted phase and, if considered as proxy for the transition temperature, aligns with the experimental trends of a lower $T_c$ under tensile (001) strain.

We then explore the effect of varying the $a$ and $c$ lattice parameters independently, observing that the M1 phase is energetically favored at large $c$, and at both large and small $a$. We analyze the changes in the nearest-neighbor V--V distances and the R phase electronic structure, and discuss several factors that can favor or disfavor the Peierls-distorted phase. While the M1 phase stabilization at large $c$ depends on an interplay between the $d_{x^2 - y^2}$ orbital filling, the $c$ direction hopping, and the stiffness of the underlying lattice, our results indicate that the former two effects are weak, and it is the structural stiffness that is the decisive factor. Additionally, the fact that we observe an effect of the basal plane strain at all, highlights the multidimensional character of VO$_2$.

Finally, we also investigate a symmetry-breaking case corresponding to an epitaxial constraint along $a$ and $c$, resulting in $a \neq b$. We note that we do not observe large differences compared to the tetragonal strain case. In particular, we confirm that this regime is also mainly dominated by the strain in the $c$ direction (now oriented in-plane).

From our analysis of the three different strain scenarios, we can thus conclude that the strain along the dimerization axis $c$ indeed plays the central role in tuning the relative stability of the R and M1 phases. While strain that changes $a$ also affects the phase stability, its main effect is mediated through the resulting change in $c$, particularly in the epitaxial case. Altogether, our results highlight that although the basal plane strain also seems to affect the structural dimerization to some extent, hinting at the three-dimensional character of VO$_2$, both structurally and electronically, it is the $c$ axis strain, whether applied directly or indirectly, that plays the dominant role in tuning the MIT in VO$_2$. Strain engineering strategies should thus target the dimerization direction of the material, even at the cost of imposing nonessential basal plane strains.

%%%%%%%%%%%%%%%%%%%%%%%%%%%%%%%%%%%%%%%%%%%%%%%%%%%%%%%%%%%%%%%%%%%%%%%%%%%%%%%%

\section*{Acknowledgments}
The authors thank A. M. Ionescu and his group for valuable discussions. This work was supported by ETH Z\"{u}rich and the Swiss National Science Foundation (Grant No.~209454). Calculations were performed on the ETH Z\"{u}rich Euler cluster.

\bibliography{strain}

\end{document}